\documentclass[12pt]{article}

\newcommand\pubnumber{LTH 492}
\newcommand\pubdate{\today}
\newcommand\hepnumber{hep-ph/0101159}

\def\csumb{Department of Mathematical Sciences\\
University of Liverpool, Liverpool L69 3BX, U.K.}
\def\support{\footnote{Work supported by the Leverhulme 
Trust and by PPARC}} 

\def\Title#1{\begin{center} {\Large\bf #1 } \end{center}}
\def\Author#1{\begin{center}{ \sc #1} \end{center}}
\def\Address#1{\begin{center}{ \it #1} \end{center}}

\newcommand\pubblock{\rightline{\begin{tabular}{l} \pubnumber\\
         \pubdate\\ \hepnumber \end{tabular}}}
\newenvironment{Abstract}{\begin{quotation}  }{\end{quotation}}
\newenvironment{Presented}{\begin{quotation} \begin{center} 
             Presented at the\end{center}
      \begin{center}\begin{large}}{\end{large}\end{center} \end{quotation}}
\def\Acknowledgments{\bigskip  \bigskip \begin{center}
          \large\bf Acknowledgements\end{center}}

\makeatletter
\def\section{\@startsection{section}{0}{\z@}{5.5ex plus .5ex minus
 1.5ex}{2.3ex plus .2ex}{\large\bf}}
\def\subsection{\@startsection{subsection}{1}{\z@}{3.5ex plus .5ex minus
 1.5ex}{1.3ex plus .2ex}{\normalsize\bf}}
\def\subsubsection{\@startsection{subsubsection}{2}{\z@}{-3.5ex plus
-1ex minus  -.2ex}{2.3ex plus .2ex}{\normalsize\sl}}

\renewcommand{\@makecaption}[2]{%
   \vskip 10pt
   \setbox\@tempboxa\hbox{\small #1: #2}
   \ifdim \wd\@tempboxa >\hsize     
       \small #1: #2\par          
     \else                        
       \hbox to\hsize{\hfil\box\@tempboxa\hfil}
   \fi}

 \def\citenum#1{{\def\@cite##1##2{##1}\cite{#1}}}
 
\newcount\@tempcntc
\def\@citex[#1]#2{\if@filesw\immediate\write\@auxout{\string\citation{#2}}\fi
  \@tempcnta\z@\@tempcntb\m@ne\def\@citea{}\@cite{\@for\@citeb:=#2\do
    {\@ifundefined
       {b@\@citeb}{\@citeo\@tempcntb\m@ne\@citea\def\@citea{,}{\bf ?}\@warning
       {Citation `\@citeb' on page \thepage \space undefined}}%
    {\setbox\z@\hbox{\global\@tempcntc0\csname b@\@citeb\endcsname\relax}%
     \ifnum\@tempcntc=\z@ \@citeo\@tempcntb\m@ne
       \@citea\def\@citea{,}\hbox{\csname b@\@citeb\endcsname}%
     \else
      \advance\@tempcntb\@ne
      \ifnum\@tempcntb=\@tempcntc
      \else\advance\@tempcntb\m@ne\@citeo
      \@tempcnta\@tempcntc\@tempcntb\@tempcntc\fi\fi}}\@citeo}{#1}}
\def\@citeo{\ifnum\@tempcnta>\@tempcntb\else\@citea\def\@citea{,}%
  \ifnum\@tempcnta=\@tempcntb\the\@tempcnta\else
  {\advance\@tempcnta\@ne\ifnum\@tempcnta=\@tempcntb \else\def\@citea{--}\fi
    \advance\@tempcnta\m@ne\the\@tempcnta\@citea\the\@tempcntb}\fi\fi}
\makeatother

\def\ttil{\tilde t}
\def\btil{\tilde b}
\def\tautil{\tilde \tau}
\def\util{\tilde u}
\def\dtil{\tilde d}
\def\etil{\tilde e}

\def\nutil{\tilde \nu} 
\def\chitil{\tilde \chi} 
\def\TeV{{\rm TeV}}
\def\GeV{{\rm GeV}}
\def\Rcal{{\cal R}}

\def\abar{{\overline a}}
\def\thetabar{{\overline \theta}}
\def\kappabar{{\overline \kappa}}

\def\qbar{{\overline q}}
\def\Tr{\, {\rm Tr}}
\def\JHEP{{\sl JHEP\ }}
\def\plb{{\sl Phys.\ Lett.\ }}
\def\npb{{\sl Nucl.\ Phys.\ }}

\def\ptp{{\sl Prog.\ Th.\ Phys.\ }}

\def\mbar{\overline{m}}
\def\thetabar{{\overline \theta}}

\def\nn{\nonumber\\}
\def\mhat{\hat{m}}
\def\semi{;\hfil\break}
\def\sy{supersymmetry}

\def\beq{\begin{equation}}
\def\eeqn{\end{equation}}
 
\newenvironment{Eqnarray}%
   {\arraycolsep 0.14em\begin{eqnarray}}{\end{eqnarray}}
\def\beqa{\begin{Eqnarray}}
\def\eeqa#1{\label{#1}\end{Eqnarray}}
\def\eeqan{\end{Eqnarray}}

\begin{document}
\begin{titlepage}
\pubblock

\vfill
\def\thefootnote{\fnsymbol{footnote}}
\Title{Anomaly Mediated Supersymmetry Breaking, D-terms and 
R-symmetry} 
\vfill
\Author{D.R.T. Jones\support}
\Address{\csumb}
\vfill
\begin{Abstract}
We explore two distinct resolutions to the tachyonic slepton 
puzzle in the Anomaly Mediated Supersymmetry Breaking scenario.
Both are based on extending the MSSM by an anomaly free $U_1$ 
symmetry, and in both cases exact RG invariance is preserved.
\end{Abstract}
\vfill
\begin{Presented}
5th International Symposium on Radiative Corrections \\ 
(RADCOR--2000) \\[4pt]
Carmel CA, USA, 11--15 September, 2000
\end{Presented}
\vfill
\end{titlepage}
\def\thefootnote{\arabic{footnote}}
\setcounter{footnote}{0}

\section{Introduction}
Recently there has been interest in a 
specific and  predictive  framework for the origin of soft supersymmetry 
breaking within the MSSM, known as 
Anomaly Mediated Supersymmetry Breaking (AMSB). 
The basic AMSB solution is 
given by\cite{amsbrefs}:
\beqa\label{amsb} M &= M_0{\beta_g\over g}\nn
h^{ijk}&=-M_0\beta_Y^{ijk}\nn
(m^2)^i{}_j &= \frac{1}{2}|M_0|^2\mu\frac{d\gamma^i{}_j}{d\mu}.
\eeqan
where $M$ is the gaugino mass, $h^{ijk}$  the $\phi^3$ coupling 
$Y^{ijk}$  the superpotential Yukawa  coupling
 and $(m^2)^i{}_j$ the  $\phi\phi^*$-mass. They are all given in terms  of
the gravitino mass, $M_0$, and the RG functions $\beta_g$ and $\gamma^i{}_j$
of the unbroken theory. It is interesting that two of these relations 
were first developed in an attempt to construct RG trajectories\cite{jjp}; 
the results for $M,h,m^2$ satisfy exactly the formulae for the 
corresponding $\beta$-functions given, for example in \cite{jjp}).

Direct application of the AMSB solution  to the MSSM leads,
unfortunately,  to negative $(\hbox{mass})^2$ sleptons: in other words,
to a theory  without a vacuum  preserving the $U_1$ of electromagnetism.
We explore two distinct  resolutions\cite{jjfid}\cite{jjrsym} 
of this dilemma, both based on
generalising the AMSB solution, while retaining the crucial property of
RG invariance,  which makes the low energy theory insensitive to the
nature of new  physics at high scales. (For some other approaches see 
Refs.~\cite{appp}-\cite{nahk}). Both our ideas are based on
extending the  MSSM with an extra $U_1$; in the second case this $U_1$ 
being associated with an $R$-symmetry. In both cases it transpires  that
requiring RG invariance of the generalised AMSB solution  means that the
$U_1$ must have no mixed anomalies with the MSSM  gauge group. 
Also in both cases, a distinguishing feature of the results 
is the existence of sum rules for the sparticle masses. 

\section{Fayet-Iliopoulos D-terms}

A modification of the AMSB solution which has been studied in some detail 
is the simple replacement $m^2\to \mhat^2$ where 
\beq
(\mhat^2)^i{}_j = (m^2)^i{}_j + m_0^2\delta^i{}_j.
\eeqn
Here $m^2$ is the basic AMSB solution form Eq.~(\ref{amsb}) and 
$m_0^2$ is constant. This is not RG invariant (for constant $m_0^2$), but if instead we have   
\beq\label{newmass}
(\mhat^2)^i{}_j = (m^2)^i{}_j + m_0^2\sum_{a=1}^{\cal N} k_a (Y_a)^i{}_j
\eeqn
then $\mhat^2$ {\it is} RG invariant, as long as each $Y_a$ corresponds 
to a $U_1$ invariance of the superpotential $W$ and also has  vanishing
mixed anomaly with each MSSM gauge group factor. This apparent miracle
occurs because  in fact the modification to $m^2$ proposed in
Eq.~(\ref{newmass}) is precisely that introduced by a set of 
Fayet-Iliopoulos (FI) $D$-terms. 

In the MSSM, there is a non-zero  FI-term, but this cannot alone solve the
slepton problem because  its $(\hbox{mass})^2$ contributions to the  LH
and RH sleptons  have opposite signs,  being dictated by the hypercharge
of the relevant field. The minimal solution we proposed was the
introduction of  a single extra $U_1$, which we denote $U_1'$.  
The MSSM does not admit such a generation independent)  anomaly-free $U_1$ 
so we need to introduce some new fields to cancel the associated 
anomalies. However cancellation of the {\it mixed\/} anomalies 
can be achieved within the MSSM itself; the hypercharges of the 
quark and Higgs multiplets are determined in terms of the lepton 
hypercharges, so that the MSSM admits two independent mixed-anomaly-free 
$U_1$ groups, the existing $U_1^Y$ and another (which could be chosen to be 
$U_1^{B-L}$\cite{nahk}).  If we also require absence of 
$(U_1')^3$ and $U'_1$-gravitational anomalies this can be achieved by 
introducing a set of MSSM singlets with $U_1'$ charges $s_i$ 
and imposing the constraints\cite{cham}
\beqa\label{dioph}
\sum_{i=1}^N s_i &= -3(2Y'_L + Y'_{\tau^c}),\quad\hbox{and}\nn
\sum_{i=1}^N s_i^3 &= -3(2Y'_L + Y'_{\tau^c})^3,
\eeqan
where the existing MSSM $U_1$ corresponds of course to 
$s_i = 2Y'_L + Y'_{\tau^c}=0$. 

The classification of rational solutions to Eq.~(\ref{dioph}) is an interesting 
Diophantine problem; but as explained in \cite{jjfid}, 
all we require for the RG invariance of  Eq.~(\ref{newmass}) is the 
existence of the sets of charges $Y_a$; there need be no 
relic of the associated gauge symmetry (or the singlets $S_i$) 
in the low energy theory.  
This is the point of view we will take from now on.

\def\tableline{\noalign{
\hrule height.7pt depth0pt\vskip3pt}}

\begin{table}[t!]
\caption{\label{Hyperchargesa}Table of $U_1$ and $U'_1$ hypercharges.}
\begin{center}
\setlength{\tabcolsep}{9pt}
\renewcommand{\arraystretch}{1.2}
\begin{tabular}{lccccccccc}
\tableline
& $Q$ & $L$ & $t^c$ & $b^c$ & $\tau^c$ & $H_1$ & $H_2$ & $S_i$ \\ \hline
& & & & & & & & \\ 
$Y$ & $\frac{1}{6}$ & ${-\frac{1}{2}}$ & $-\frac{2}{3}$ & $\frac{1}{3}$ &  
${1}$ & $-\frac{1}{2}$ &  $\frac{1}{2}$ & 0 \\ 
& & & & & & & & \\
\hline
& & & & & & & & \\
$Y'$  & $\frac{7}{3}$ & ${-7}$ & $\frac{5}{3}$ & $-\frac{19}{3}$
& ${3}$ & $4$ & $-4$ & $s_i$
\\ 
& & & & & & & & \\
\hline
\end{tabular}
\end{center}
\end{table}

In Table~(\ref{Hyperchargesa}) we give a possible set of $U_1'$ charges with 
the $U_1^Y$ ones for comparison. 
This set of $Y'$ charges correspond to requiring $\Tr (YY') = 0$;
of course the result is a linear combination of $U_1^Y$ and 
$U_1^{B-L}$. The outcome is that the squark and slepton masses
are given by 
\beqa
\mbar^2_Q & = & m^2_Q +\frac{1}{6}\zeta_1 +\zeta_2 Y'_Q,\quad
\mbar^2_{t^c} = m^2_{t^c} -\frac{2}{3}\zeta_1 + \zeta_2 Y'_{t^c},\nn
\mbar^2_{b^c} & = & m^2_{b^c} +\frac{1}{3}\zeta_1  +\zeta_2 Y'_{b^c},\quad
\mbar^2_L =  m^2_L -\frac{1}{2}\zeta_1 + \zeta_2 Y'_L,\nn
\mbar^2_{\tau^c} & = & m^2_{\tau^c} +\zeta_1 + \zeta_2 Y'_{\tau^c}
\eeqan
where dependence on the FI coefficients, 
the Higgs vevs and the singlet sector
is subsumed into the parameters $\zeta_{1,2}$. There is a substantial 
region of the $\zeta_{1,2}$-plane such that all the above $\mbar^2$ parameters 
are positive (the precise region depending on the input $\tan\beta$).
The result is interesting phenomenology. The gaugino spectrum 
is the same as in previous AMSB scenarios, with a near degenerate wino 
and neutralino, the latter being (in some regions of parameter space) 
the LSP.

For a choice of $m_0=40\TeV$, $\tan \beta=5$, $\zeta_1=0.2$,
 $\zeta_2=-.02$, we find $|\mu_s| = 645\GeV$ and (choosing $\mu_s > 0$) a 
mass spectrum given by:
\beqa
m_{\ttil_1}& = &575, \quad m_{\ttil_2}=861, \quad m_{\btil_1}=825, \quad   
m_{\btil_2}=1040,
\quad m_{\tautil_1}=137,
\quad m_{\tautil_2}=339,\nn
m_{\util_L}& = &931, \quad m_{\util_R}=851,\quad m_{\dtil_L}=935,
\quad m_{\dtil_R}=1045,
\quad m_{\etil_L}=139,\quad m_{\etil_R}=339,\nn
m_{\nutil} & = & 112,\quad m_{h,H} = 110, 455, \quad 
m_A=453, \quad m_{H^{\pm}} = 461,
\quad m_{\tilde\chi_{1,2}^{\pm}} = 104, 649\nn
m_{\chitil_{1\cdots4}} & = & 103, 366, 648, 658,\quad m_{\tilde g} = 1007,
\eeqan
where all masses are given in GeV.

A characteristic feature of the present setup  is 
the existence of sparticle mass sum rules such as 
\beq\label{sumfi}
m_{\ttil_1}^2+m_{\ttil_2}^2+m_{\btil_1}^2+m_{\btil_2}^2 
- 2(m_t^2 + m_b^2) = 2.75m^2_{\tilde g}.
\eeqn 
The numerical coefficient here depends only on the input $\tan\beta$ 
(here taken to be $\tan\beta = 5$), and also 
(weakly) on $m_0$, here taken to be $40\TeV$. 

\section{R-symmetry and Yukawa textures}

There is an alternative generalisation of the AMSB solution for $m^2$ 
from Eq.~(\ref{amsb}) as follows:
\beq\label{mnew} 
(m^2)^i{}_j = \frac{1}{2}|m_0|^2\mu\frac{d\gamma^i{}_j}{d\mu}
+ \mbar_0^2 (\gamma^i{}_j + \qbar^i \delta^i{}_j).
\eeqn
This is also RG invariant to all orders as long as $\qbar^i$ satisfy 
the following constraints:

\beqa\label{qrels}
(\qbar^i+\qbar^j+\qbar^k)Y_{ijk} & = & 0 \nn
2\Tr\left[\qbar C(R)\right] + Q & = & 0, 
\eeqan
where $Q$ is the one loop $\beta_g$ coefficient and $C(R)$ is the 
quadratic matter Casimir.  It is easy to show\cite{appp}\cite{lura}  
that Eq.~(\ref{qrels}) corresponds precisely to requiring 
that the theory have a non-anomalous  
${\cal R}$-symmetry (which we denote ${\cal R}$, to 
avoid confusion with our notation $R$ for group representations).  
If we set 
\beq\label{qcharg}
\qbar^i = 1 -\frac{3}{2}r^i,
\eeqn
then we see that Eq.~(\ref{qrels}) 
corresponds to $(r^i+r^j+r^k)Y_{ijk} = 2Y_{ijk}$, 
which is the conventional ${\cal R}$-charge normalisation.

Turning to the MSSM we find that,
as in the previous section, our solution will retain the crucial RG 
invariance as long as all the {\it mixed\/} anomalies of the 
${\cal R}$-symmetry with the MSSM gauge group vanish.  
The MSSM does not admit such a generation independent ${\cal R}$-symmetry;
however it {\it does\/} admit one that permits only 3rd generation 
Yukawa couplings and has identical ${\cal R}$-charges for the 
first two generations.  We find that this can be achieved for 
for arbitrary values of the leptonic charges with the quark and Higgs charges 
determined as follows (we work with the fermionic charges,
related to the $\Rcal$-charges by $q_f = r -1$):

\beqa\label{ansol}
q_3 & = & \frac{4}{9}-\frac{1}{3}l_3 -\frac{1}{9}\frac{\kappabar}{\kappa}
\nn
u_3 & = & -\frac{22}{9} -\frac{2}{3}l_3 - e_3 
+\frac{1}{9}\frac{\kappabar}{\kappa}
\nn
d_3 & = & -\frac{4}{9} +\frac{4}{3}l_3 + e_3 
+\frac{1}{9}\frac{\kappabar}{\kappa}
\nn
q_1 & = & -\frac{101}{90} -\frac{1}{3}\kappa +\frac{1}{15}l_3 +\frac{1}{5}e_3 
+\frac{1}{30}\kappabar  +\frac{1}{18}\frac{\kappabar}{\kappa}
\nn
u_1 & = & -\frac{79}{90} -\frac{2}{3}\kappa -\frac{16}{15}l_3 
-\frac{6}{5}e_3 
-\frac{1}{30}\kappabar - \frac{1}{18}\frac{\kappabar}{\kappa}
\nn
d_1 & = & \frac{101}{90} +\frac{4}{3}\kappa +\frac{14}{15}l_3 
+\frac{4}{5}e_3-\frac{1}{30}\kappabar 
- \frac{1}{18}\frac{\kappabar}{\kappa}\nn
h_2 & = & - h_1 = l_3+e_3 +1, 
\eeqan
where $\kappa = l_1-l_3+e_1-e_3 - 3$, and 
$\kappabar= -12l_3-16e_3+10e_1 -23$. As explained above, 
we have imposed $q_1 = q_2$ etc.

Thus for any set of rational values for the leptonic charges there 
exist rational values for all the charges. it is clear therefore that we 
can potentially resolve the tachyonic slepton problem, since we can 
choose the lepton ${\cal R}$-charges so that all the $\qbar$ contributions  
to Eq.~(\ref{mnew}) are positive for the sleptons. Of course we will need 
to check that the corresponding contributions for the squarks and Higgses 
do not cause problems. 

We can get a different handle on the 
$\Rcal$-charge assignments by relating them to  
a possible origin of the light quark and lepton masses. 
Suppose\cite{cdfr}
there are higher-dimension terms in the effective field 
theory of the form (for the up-type quarks)
$H_2 Q_i u^c_j (\frac{\theta}{M_U})^{a_{ij}}$ or
$H_2 Q_i u^c_j (\frac{\thetabar}{M_U})^{\abar_{ij}}$, 
where $\theta,\thetabar$ is a pair of 
MSSM singlet fields with $\Rcal$-charges $\pm r_{\theta}$ 
that get equal vacuum expectation values, 
and $M_U$ represents some high energy new physics scale (with similar 
terms for the light down quarks and leptons). Evidently the $\Rcal$-charge 
assignments will then dictate the texture of the Yukawa couplings, 
via the relation $h_2 +q_1 +u_1 +a_{11} r_{\theta}= -1$ and similar identities.
if we suppose identical textures for the down quarks and leptons 
then we find 
\beq\label{dl}
\kappa=-\frac{3}{2}, \qquad \kappabar=-\frac{21}{2}-\frac{9}{4}\lambda,
\eeqn
where $\lambda=2l_3+e_3$. The only value of $\lambda$ we have
found which leads to nice textures with only one pair of $\theta$, 
$\thetabar$ fields is $\lambda=-\frac{1}{3}$, which leads to
the set of fermionic $\Rcal$-charges shown in Table~(\ref{Hyperchargesb}). 
\begin{table}[t!]
\caption{\label{Hyperchargesb}The fermionic $\Rcal$-charges 
for the case $\Delta_d=\Delta_L$}
\begin{center}
\setlength{\tabcolsep}{9pt}
\renewcommand{\arraystretch}{1.2}
\begin{tabular}{cccccc}
\tableline
$ q_3      $ & $ l_3           $ & $ u_3     $ & $ d_3 $ & 
$ e_3 $ & $ q_1$\\ \tableline
& & & & & \\ 
  $ \frac{e}{6} - \frac{2}{9}
  $ & $-\frac{e}{2}-\frac{1}{6}$ & $-\frac{2e}{3}-\frac{29}{18}
  $ & $\frac{e}{3}+\frac{1}{18}  $ & $ e $ & $  \frac{e}{6}-\frac{43}{72}$
\\ & & & & & \\ \tableline
\end{tabular}

\vskip1em

\setlength{\tabcolsep}{9pt}
\renewcommand{\arraystretch}{1.2}
\begin{tabular}{cccccc}
\tableline
$l_1           $ & $ u_1     $ & $ d_1 $ & $ e_1
$ & $ H_1      $ & $ H_2$\\ 
\tableline & & & & & \\   
$ -\frac{e}{2}+\frac{5}{24}
  $ & $ -\frac{2e}{3}+\frac{19}{72}        $ & $ \frac{e}{3}-\frac{77}{72}
  $ & $e+\frac{9}{8}            
  $ & $ -\frac{e}{2} -\frac{5}{6}$ & $ \frac{e}{2} +\frac{5}{6}$\\ 
& & & & & \\ \tableline
\end{tabular}
\end{center}
\end{table}

With this charge assignment we find, (for arbitrary $e$) but setting 
$r_{\theta}=\frac{3}{8}$,  
Yukawa textures of the form 
\beq\label{textureq}
\Delta_u =
\pmatrix{\epsilon^4
& \epsilon^4 & \epsilon\cr
\epsilon^4
& \epsilon^4 & \epsilon\cr
\epsilon^5
& \epsilon^5 & 1\cr},\quad
\Delta_d = \Delta_L=
\pmatrix{\epsilon^4
& \epsilon^4 & \epsilon\cr
\epsilon^4
& \epsilon^4 & \epsilon\cr
\epsilon^3
& \epsilon^3 & 1\cr}
\eeqn
The quark
and lepton mass hierarchies and the CKM matrix can  be produced with
matrices of these generic structures, see \cite{jjrsym}. The 
phenomenology of Flavour Changing Neutral Currents (in both 
hadronic and leptonic sectors) and CP-violation
effects clearly deserve a detailed investigation. 

It is easy to show that  as long as $-\frac{1}{3} < e < \frac{1}{3}$ and
$\mbar_0^2<0$, the contribution to each slepton mass term due to the
$\qbar$ term in  Eq.~(\ref{mnew}) will be positive, and we may expect to
achieve a viable spectrum; however, it turns out that it is still
non-trivial to obtain   an acceptable minimum because, for example, if
$e=0$ and $\mbar_0^2<0$, the $\mbar_0^2\qbar$ contributions to
Eq.~(\ref{mnew})  from $u_3$, $q_1$ and $d_1$ are negative. We find
in fact that we need to have $e < 0$.  

A variety of mass spectra for $m_0=40\TeV$ (corresponding to a gluino
mass of around $1\TeV$), but with different values of $\tan\beta$, $e$
and $\mbar^2_0$, is  presented in table~\ref{spmasses}; we were unable
to find any values  of $e$ and $\mbar^2_0$ corresponding to an
acceptable spectrum for  $\tan\beta$ significantly larger than $10$.  
The heaviest sparticle masses scale with $m_0$ and are given roughly by
$M_{\rm SUSY} = \frac{1}{40}m_0$. A characteristic feature of AMSB
models is the near-degenerate light  charged and neutral winos;  this
prediction, as in the FI case,  is preserved in the scenario presented
here.   The main distinction from the FI case is the large splitting
between  the third generation and the other two, caused by the
generation-dependent  $\Rcal$-charge assignments. Moreover, unusual  is
the possibility  (exemplified in the first three columns of
table~\ref{spmasses}) that the  $\nutil_{\tau}$ is the LSP.  As is well
known, radiative corrections   give a sizeable upward contribution  to
the mass of the light CP-even Higgs,  and so we have included the
one-loop calculation.

\begin{table}[t!]
\caption{\label{spmasses}The sparticle masses (given in $\GeV$) 
for the $U_1^R$ case}
\begin{center}
\setlength{\tabcolsep}{9pt}
\renewcommand{\arraystretch}{1.2}
\begin{tabular}{cccccc}
\tableline
$\tan\beta (\hbox{sign}~\mu_s) $ & $ 2(+)  $ & $2(-)  $ & $ 5(+)  $ & $5(+) $ 
& $10(+)$ 
\\ \tableline 
$e$ & $ -1/9 $ & $ -1/9  $ & $ -1/9  $ & $ -2/9 $ & $-2/9$ \\ \tableline
$\mbar_0^2 (\TeV^2)$ & $  -0.1$ & $-0.1  $ & $ -0.1  $ & $  -0.25$ & 
$-0.2$ \\ \tableline
$ \ttil_{1,2} $ & $ 652,882 $ & $615,908$ & $ 567,876  $ & $302,879 $ & 
$404,875$ \\ \tableline
$\btil_{1,2} $ & $ 865, 977  $ & $865, 977 $ & $ 843,974  $ & $853,1009 $ 
& $843,987$ \\ \tableline
$\tautil_{1,2} $ & $ 94,110  $ & $ 87,116 $ & $ 75,127  $ & $ 136,289  $ & 
$86,251$ \\ \tableline
$\util_{L,R} $ & $918,997 $ & $918,997$ & $917,997  $ & $ 880,1084,  $ & 
$892,1057$ \\ \tableline
$\dtil_{L,R} $ & $920,887$ & $920,887$ & $921,887 $ & $ 884,776 $ & 
$896,814$ \\ \tableline
$\etil_{L,R} $ & $260,423$ & $260,423$ & $261,423 $ & $473,664 $ 
& $418,590$ \\ \tableline
$\nutil_{\tau}$ & $ 83$ & $83$ & $ 73 $ & $ 277 $ & $234$ \\ \tableline 
$\nutil_e $ & $ 251$ & $251$ & $ 249 $ & $467 $ & $410$ \\ \tableline   
$h $ & $ 96 $ & $ 105 $ & $119  $ & $114 $ & $124$ \\ \tableline
$H $ & $ 598 $ & $ 598 $ & $ 585  $ & $ 121 $ & $308$ \\ \tableline
$A $ & $ 593$ & $ 593$ & $584  $ & $110 $ & $307$ \\ \tableline
$H^{\pm} $ & $ 599$ & $599$ & $ 590 $ & $137 $ & $318$ \\ \tableline  
$\chitil^{\pm}_1 $ & $ 98 $ & $116 $ & $ 104 $ & $ 101$ & $106$ \\ \tableline
$\chitil^{\pm}_2 $ & $ 628 $ & $625 $ & $ 663 $ & $449 $ & $530$ \\ \tableline
$\chitil_1 $ & $98$ & $ 115 $ & $103 $ & $99$ & $103$ \\ \tableline
$\chitil_2 $ & $ 364 $ & $372 $ & $ 367 $ & $357$ & $365 $ \\ \tableline
$\chitil_3 $ & $ 619 $ & $ 620 $ & $662  $ & $446 $ & $532$ \\ \tableline
$\chitil_4 $ & $637$ & $ 628 $ & $ 672  $ & $470$ & $ 544 $ \\ \tableline
${\tilde g} $ & $ 1008 $ & $ 1008 $ & $1008    $ & $ 1008$ & $1008$\\
\tableline
\end{tabular}
\end{center}
\end{table}

As in the FI case,  however, a salient feature of the model 
is the existence of sum rules for the sparticle masses. 
These sum rules 
follow from Eq.~(\ref{ansol}); and thus for the particular solution 
exhibited in table~\ref{spmasses}, 
they are independent of $e$. We find for example the
following relation: 
\beq\label{sumthr}
m_{\ttil_1}^2+m_{\ttil_2}^2+m_{\btil_1}^2+m_{\btil_2}^2
- 2(m_t^2 + m_b^2)-2.75m^2_{\tilde g} =  0.92\mbar_0^2\TeV^2,
\eeqn
where we have again taken $\tan\beta = 5$ and $m_0 = 40\TeV$.

Note the similarity with the corresponding  one in the FI  scenario
described in the previous section, Eq.~(\ref{sumfi}); the distinction
lies in the non-zero  RHS in Eq.~(\ref{sumthr}), which can be traced
back to the dependence  on $\gamma$ in Eq.~(\ref{mnew}).

\section{Conclusions}

We have shown that by extending the MSSM  with  a $U_1$ or a $U_1^R$
(which may or may not be associated with a physical  vector boson), it
is possible to construct solutions to the running equations for $m^2$,
$M$ and $h$ that are  completely RG invariant, and leads to a
phenomenologically acceptable theories, resulting  in a distinctive
spectrum with  sum rules for the sparticle masses. In both cases the
additional source  of \sy-breaking may be provided by the vacuum
expectation value of a $D$-term. 

In a recent paper\cite{nahk}, it was shown how our first scenario can 
be compatible with currently fashionable braneworld scenarios, with 
breaking of both supersymmetry and the extra $U_1$ occurring on a hidden
brane;  the incorporation of massive neutrinos was also considered.  It
would be interesting to perform a similar construction  for the
$\Rcal$-symmetry case. 

\Acknowledgments

Warm thanks to Ian Jack, with whom all the work described here, and more, 
 was done.  I am also grateful  to the Leverhulme Trust and PPARC for 
financial support, and to Howie Haber and the other organisers  of
RADCOR 2000 for the opportunity to visit Carmel and  participate in this
meeting.


\begin{thebibliography}{99}



\bibitem{amsbrefs}
L. Randall and R. Sundrum,  \npb {\bf B557} (1999) 79\semi
G.F. Giudice, M.A. Luty, H. Murayama and  R. Rattazzi,
\JHEP 9812 (1998) 27\semi
I. Jack and D.R.T.~Jones, \plb {\bf B465} (1999) 148.
\bibitem{jjp}I.~Jack, D.R.T.~Jones and A.~Pickering,  \plb {\bf B426} (1998) 73.
\bibitem{jjfid}I.~Jack and D.R.T.~Jones, \plb {\bf B482} (2000) 167.
\bibitem{jjrsym}I.~Jack and D.R.T.~Jones, \plb {\bf B491} (2000) 151.
\bibitem{appp}A. Pomarol and  R. Rattazzi, \JHEP {\bf 9905} (1999) 013.
\bibitem{lura}M.A. Luty and R. Rattazzi, \JHEP {\bf 9911} (1999) 001.
\bibitem{katz} E. Katz, Y. Shadmi and Y. Shirman, \JHEP {\bf 9908} (1999) 015.
\bibitem{chako}
Z. Chacko, M.A. Luty, I. Maksymyk and E. Ponton, \JHEP {\bf 0004} (2000) 001.
\bibitem{gherg} 
T.~Gherghetta, G.F. Giudice and  J.D. Wells, \npb {\bf B559} (1999) 27.
\bibitem{izaw}K.I.~Izawa, Y.~Nomura and T.~Yanagida, 
\ptp {\bf 102} (1999) 1181.
\bibitem{kkribs}D.E.~Kaplan and G.D.~Kribs, \JHEP {\bf 0009} (2000) 048.
\bibitem{marcela}M. Carena, K. Huitu and T. Kobayashi, 
\npb {\bf B592} (2000) 164.
\bibitem{ben}B.C. Allanach and A. Dedes, \JHEP {\bf 0006} (2000) 017.
\bibitem{nahk}N. Arkani-Hamed, D.E. Kaplan, H. Murayama and Y. Nomura, 
LBNL-47184, [hep-ph/0012103].
\bibitem{cham}A.H. Chamseddine and  H. Dreiner, \npb {\bf B447} (1995) 195.
\bibitem{chamdr}A.H. Chamseddine and H. Dreiner, 
\npb {\bf B458} (1996) 65\semi
D.J. Castano, D.Z. Freedman and C. Manuel, \npb {\bf B461} (1996) 50.
\bibitem{cdfr}C.D. Froggatt and H.B. Nielsen, \npb {\bf B147} (1979) 277.
\end{thebibliography}
\end{document}